# A centimeter-scale achromatic hybrid metalens with polarization-insensitivity in the visible

*Tie Hu, Shengqi Wang, Yunxuan Wei, Liqing Wen, Xing Feng, Ming Zhao\*, and Zhenyu Yang\**

T. Hu, S.Q.Wang, L. Q. Wen, M. Zhao, and Z. Y. Yang

School of Optical and Electronic Information, Huazhong University of Science and Technology, Wuhan 430074, China

E-mail: zyang@hust.edu.cn, zhaoming@hust.edu.cn

Y. X. Wei

Ming Hsieh Department of Electrical and Computer Engineering, University of Southern California, Los Angeles, California 90089, USA

X. Feng

Huawei Technologies Co., Ltd., Shenzhen, 518111, China

**Abstract:** Metalenses, featuring ultra-compactness and CMOS compatibility, are limited by the compromise between the diameter, numerical aperture, and working waveband. To address this problem, we propose and numerically demonstrate a centimeter-scale metasurface-refractive hybrid metalens working in the band of 440 - 700 nm. Revisiting the general Snell law, we present the phase profile of a chromatic aberration correction metasurface that can apply to a plano-convex refractive lens of an arbitrary surface type. Simulated by our semi-vector method, the designed achromatic hybrid metalens achieves 81% chromatic aberration suppression and polarization insensitivity. Broadband imaging results of the hybrid metalens are further provided, verifying the achromatism of the designed hybrid metalens. It can find applications in camera lenses and other optical systems that need compact, high-performance lenses.

1. Introduction:

Single refractive lense suffers severe chromatic aberrations due to the positive dispersion of their composed materials. To address this problem, the traditional solution is to cascade multiple lenses composed of different materials. Later, assisted with the negative dispersion property of diffractive optical elements[1], diffractive-refractive hybrid lenses were proposed as another option to mitigate the chromatic aberrations. These hybrid lenses can work at several discrete wavelengths[2]. However, all these methods are difficult to realize a broadband achromatic lens with compactness[3].

Ultra-compact metalenses, well known for flexible manipulation of the fundamental properties of the incident light, provide a promising solution. Several methods have been proposed to compensate for the chromatism of metalens[4], including linear dispersion engineering[5], annular interference[6], spatial multiplexing[7], multiple resonances[8], and computational optimization[9]. However, due to the insufficient dispersion or phase compensation provided by subwavelength meta-atoms, these methods face a trade-off between the numerical aperture, the diameter, and the working bandwidth[10]. To overcome the limitations, metasurface-refractive hybrid metalenses, combining the focusing performance of the refractive lens and flexible dispersion manipulation of the metasurface, have been demonstrated. Chen et al. realized a millimeter-scale polarization-insensitive hybrid metalens with F/# = 6.64, based on polarization-sensitive nanofins to provide first-order and high-order dispersion manipulation[11]. Sawant et al. proposed a polarization-sensitive hybrid lens by adopting a geometric phase metasurface[12]. Other approaches such as multi-level diffractive lenses[13] and phase plate-metasurface hybrid lenses[14] are also proposed. There is still a lack of a centimeter-scale polarization-insensitive achromatic lens in the visible.

Moreover, it is still a big challenge to accurately characterize large-scale meta-elements ( e.g., metalens and metasurface-refractive hybrid lens ) by numerical simulations[15]. Maxwell-equation-based vector methods, such as the finite time domain difference (FDTD) method and the finite element method, are usually employed to analyze the transmitted electromagnetic field of metasurface with a size of 100 $\lambda$-200 $\lambda$[16]. However, these methods are time-consuming and even impossible to be used to perform the simulations of large-scale ( $\geq 1000\lambda$ ) meta-elements. On the other hand, the commercial optical simulation software ( such as Zemax OpticStudio, Ansys Inc. ) mainly based on the ray tracing method, is fit for imaging tests and aberration analysis but difficult to consider complex diffraction effects, especially concerning metasurface. Efforts should be made to develop a rapid and relatively accurate simulation method able to characterize large-scale meta-elements.

Here, considering the nonparaxial effect and using the propagation phase principle, we realize a centimeter-scale aspherical-lens-based achromatic hybrid metalens with polarization insensitivity in the waveband of 440-700 nm, featuring a diameter of 1 cm, F/# = 2.64 and a chromatic aberration correction ratio of 81%. In this work, the general phase profiles of the metasurface are proposed to suppress the chromatism of a plano-convex lens of an arbitrary surface type. Second, polarization-insensitive meta-atoms are selected to provide the target phases of the metasurface, without changing the polarization of the incident light. . Besides, the semi-vector method is introduced to perform simulations of large-scale meta-elements, such as the focusing characterizations of metasurface-refractive hybrid metalenses. The imaging tests under broadband visible light are also provided in 3.2. Our work offers an effective solution for the design and simulation of the large-scale hybrid metalenses

## 2. Principle and Design

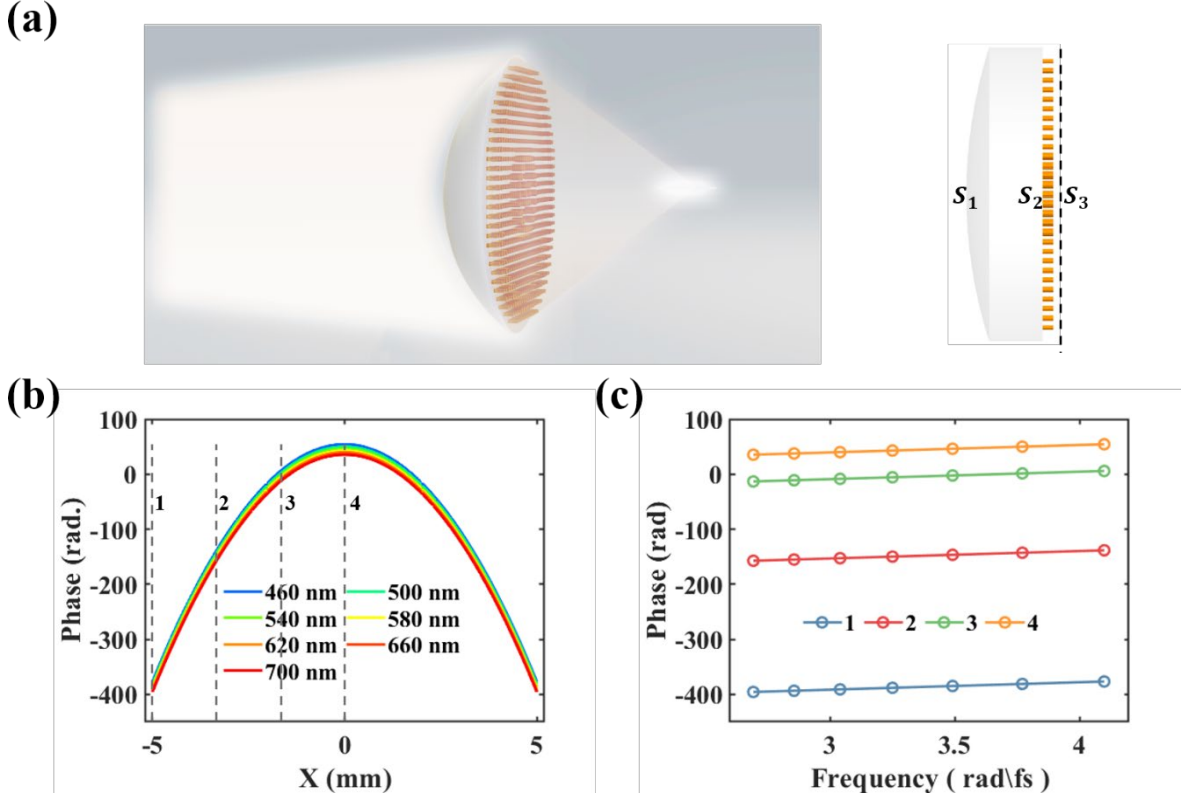

**Figure 1.** Scheme of the achromatic hybrid metalens. (a) Illustration of the hybrid metalens composed of an aspherical refractive lens and a metasurface with a diameter of 1 cm and an effective focal length of 26.4 mm. The insert is the side view of the hybrid metalens. Surfaces $S_1$, $S_2$, and $S_3$ respectively denote the front surface, the rear surface of the refractive lens, and the near-field plane after the metasurface. (b) Phase profiles of the chromatic aberration correction metasurface at wavelengths from 460 nm to 700 nm with a step of 40 nm. The black dashed lines (denoted with the orders "1", "2", "3", and "4") present the phases at the four different coordinates. (c) Detailed phases at the selected four coordinates in (b).

As shown in Figure 1 (a), the hybrid metalens consists of an aspherical lens made of S-LAH64 and a metasurface on the surface $S_2$ (shown in the insert of Figure 1(a)). The designed hybrid metalens can focus the broadband visible light to approximately the same focal plane. The positive dispersion of the aspherical lens is compensated by the negative dispersion of the metasurface, realizing a nearly chromatic-aberration-free metalens. The phase distribution of the metasurface can be derived based on the generalized Snell law and the model of the truncated waveguide-like meta-atom. The key is to design a metasurface with a constant group delay while satisfying the desired phase distributions. The phase profile of the metasurface (detailed deductions can be seen in Supplement 1) is described by Eq. (1).

$$\varphi(r_i, \omega) = L(r_i) + c_2 \omega + c_1, \qquad (1)$$

where $\omega$ is the optical frequency of the incident light, $r_i$ is the height of incident light, $c_2$ is a constant, $c_1$ is a constant reference phase. $c_1$ and $c_2$ can be optimized to best match the meta-atom library with the minimal average phase error. $L(r_i) = \sum_m A_m r_i^m$ is the wavelength-independent phase only related to $r_i$, $A_m$ is the polynomial coefficient of phase related to the shape and material of the refractive lens. Derivation of optical frequency on both sides of Eq. (1), the group delay of the metasurface is

$$\frac{d\varphi}{d\omega} = c_2. \qquad (2)$$

Eq. (2) means that there is a constant group delay distribution across the chromatic aberration correction metasurface. If $c_2 = 0$, the phases will be the same for all incident wavelengths and only the wavelength-independent geometric phase can fulfill these target phases. If $c_2 \neq 0$, the phase of the metasurface will linearly change with the optical frequency and the wavelength-dependent phases should be achieved by carefully selecting those meta-atoms with the same group delay.

Figure 1 (b) shows the target phase profiles along the x-axis at different wavelengths. For visibility, the detailed phases of the selected four coordinates on the metasurface are displayed in Figure 1 (c), the slopes of these lines are 13.54 fs.

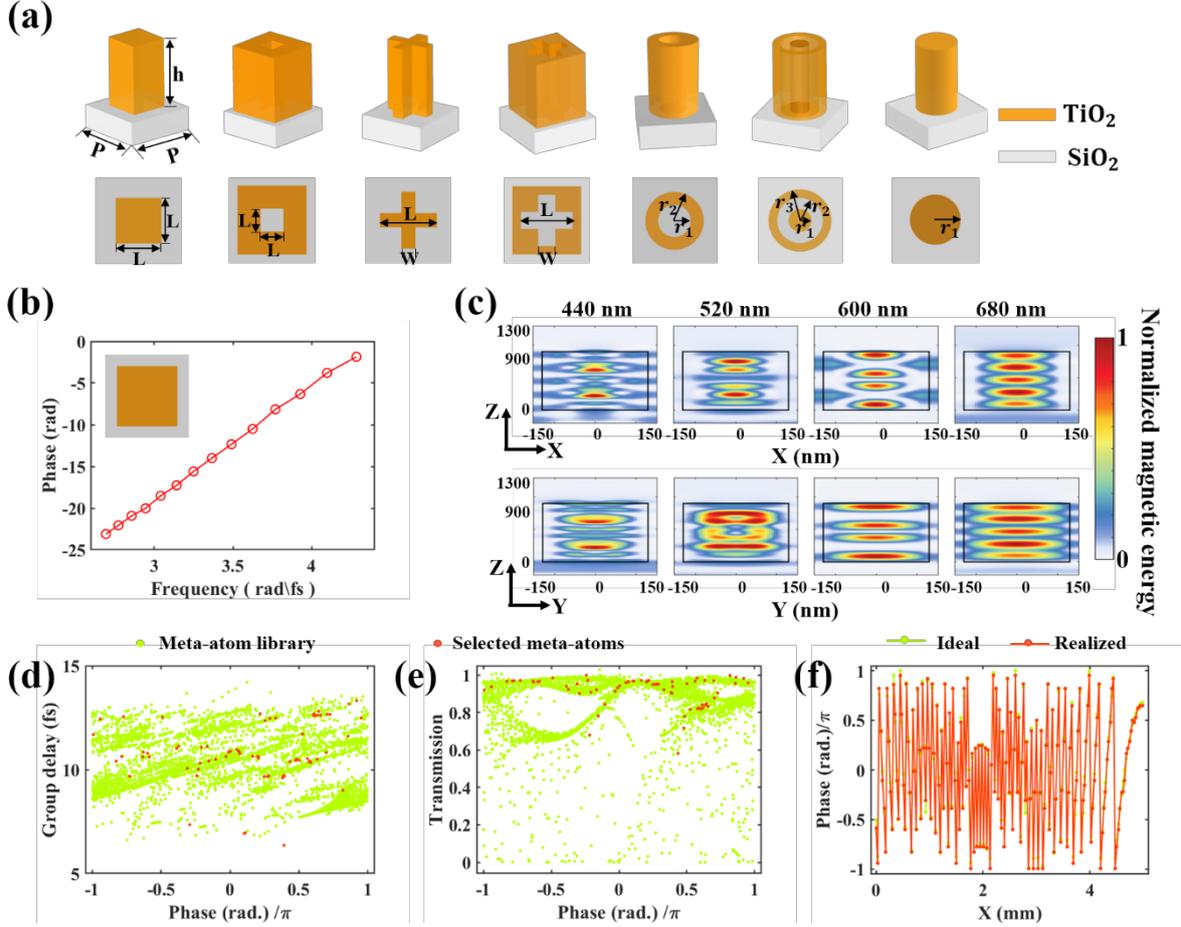

**Figure 2.** The design of the meta-atom library. (a) Illustration of the seven types of meta-atoms. (b) The phase shifts of a cross meta-atom. The length L is 216 nm, and the width W is 212 nm. The group delay is 13.66 fs and the R-square coefficient is 0.99 calculated by a quadratic polynomial fitting. The insert is the top view of this meta-atom. (c) The normalized magnetic energy density profiles of the x-z and y-z cross plane. The black lines denote the outline of this meta-atom shown in (b). (d-f) The library data and the selected data at the wavelength of 540 nm. (d) The phase-group delay map, (e) the phase-transmission map, and (f) the phase along with the radial coordinate. Yellowish green and brown markers present the library data and the selected data, respectively.

To achieve the phase profiles mentioned above for the visible wavelengths, $TiO_2$ and $SiO_2$ are selected to compose the nanopillar and the basement for the meta-atom, respectively. As illustrated in Figure 2(a), a total of seven kinds of meta-atoms are designed and optimized to comprise the metasurfaces, with an identical height of 900 nm, period of 300 nm, and polarization insensitivity due to the four-fold symmetry. Figure 2(b) shows the phase shift of the cross meta-atom with the wavelengths. This curve can be fitted by a quadratic polynomial, with an R-square coefficient of 0.99 and a group delay of 13.66 fs. As depicted in

Figure 2(c), there are clear waveguide effects of the meta-atom, which verifies that the fields are well confined in the dielectric nanopillar. The phase shifts and fields of other meta-atoms can be found in Supplement 2. Through the FDTD method, the optical responses of the meta-atom library are simulated and collected in Figure S4 in Supplement 3.

To select the optimal meta-atom at each coordinate $(x, y)$ of a metasurface, we devise a customer global optimization algorithm to find the best combination of $c_1$ and $c_2$ ( detailed method can be seen in Supplement 4 ). The core of the optimization method is to minimize the average phase error between the ideal phase profile and the simulated phase shift stored in the meta-atom library across all designed wavelengths and coordinates. As shown in Figures 2(d) and (f), the selected phases of the metasurface fit well with the ideal target phases at the wavelength of 540 nm although there is a slight difference in group delay distribution as a result of an insufficient meta-atom library. Figure 2(e) shows that most of the selected transmissions exceed 0.8 at the wavelength of 540 nm.

Once the optimization is accomplished, the metasurface is fully established and the near-field distribution $A_M e^{i\varphi_M}$ can be obtained from the optical responses of the selected meta-atoms. In this text, as the metasurface is assumed to be fabricated on the planar substrate of the aspherical lens and their optical axes are aligned to each other, the output field $A_H e^{i\varphi_H}$ at the surface $S_3$ is

$$A_H = A_M \times A_A \tag{3}$$
$$\varphi_H = \varphi_M + \varphi_A, \tag{4}$$

where $A_H$ and $\varphi_H$ are respectively the amplitude and phase responses of the hybrid metalens, $A_A$ and $\varphi_A$ are respectively the transmitted amplitude and phase distributions at the plane $S_2$. Since commercial refractive lenses are coated with antireflection films, $A_A$ approximates unity.

Finally, the Rayleigh Sommerfeld vector diffraction algorithm is used to calculate the focusing fields of the hybrid metalens based on the near field $A_H e^{i\varphi_H}$. Table 1 summarizes some merits used in this paper.

**Table 1**  Definitions of lens's merits

| maximal focal length shift | chromatic aberration correction ratio | focusing efficiency |
|---|---|---|
| $\Delta f = f_{max} - f_{min}$ | $(1 - \Delta f_{hybrid}/\Delta f_{bare}) \times 100\%$ | $(P_{Airy}/P_{inc}) \times 100\%$ |

Notes: $f_{max}$ and $f_{min}$ are the maximal and minimal focal lengths, $\Delta f_{hybrid}$ and $\Delta f_{bare}$ are respectively the maximal focal length shift of the bare lens and the hybrid metalens, $P_{Airy}$ is the power occupied by the focal spot with a radius of the ideal Airy radius, $P_{inc}$ is the transmitted power.

## 3. Results and Discussion
### 3.1 Focusing with the hybrid metalens

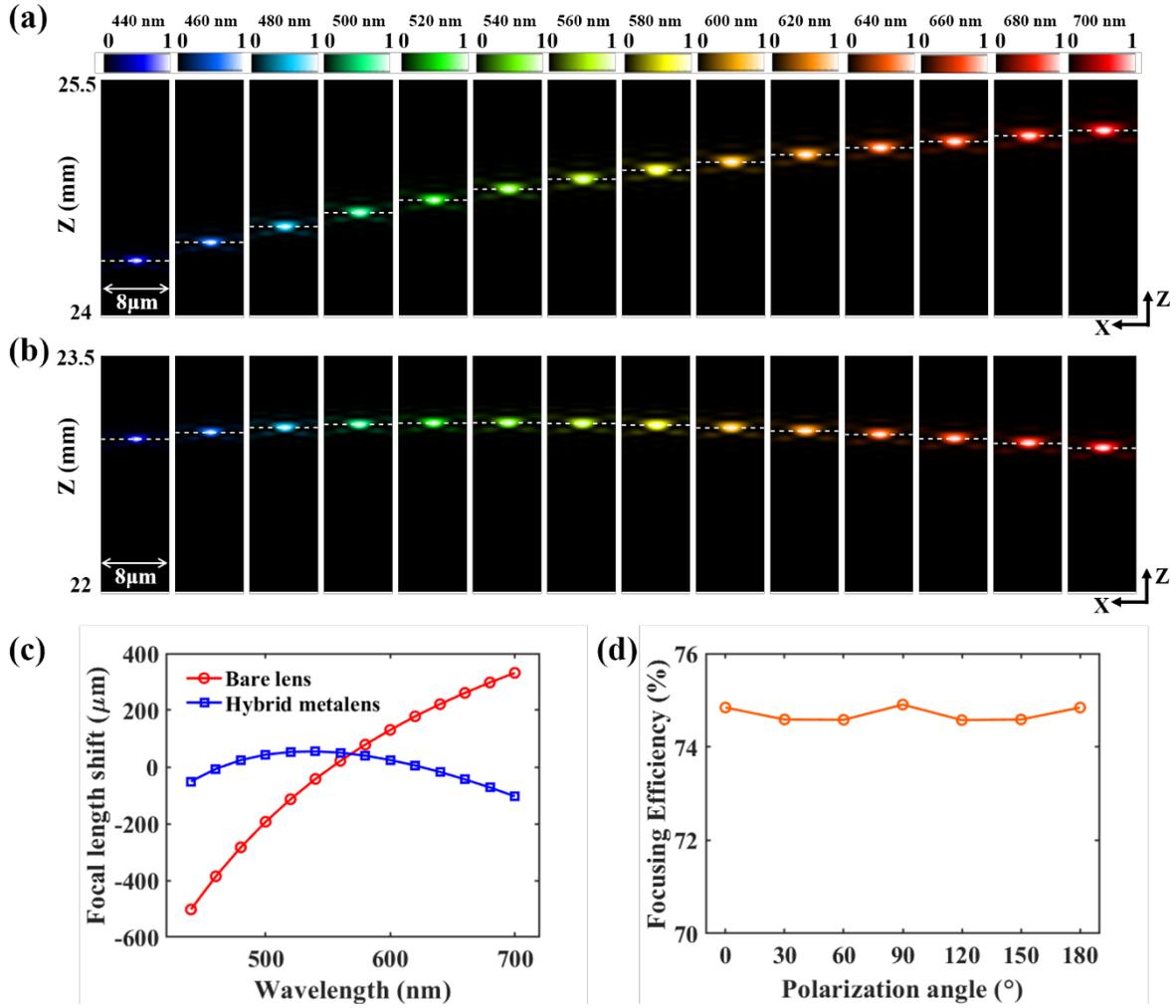

**Figure 3.** The achromatic focusing behavior of the hybrid metalens. (a, b) The normalized intensity distributions of the x-z plane at the wavelengths from 440 nm to 700 nm with a step of 20 nm. (a) Results of the bare lens. (b) Results of the hybrid metalens. White dashed lines denote the focal planes. (c) The focal length shifts versus different wavelengths. The blue squares and red circles represent the results of the bare lens and the hybrid metalens, respectively. (d) The focusing efficiency versus the incident polarizations at the wavelength of 640 nm.

Figures 3(a) and 3(b) show the normalized intensity distributions at the x-z plane for the bare aspherical lens and the hybrid metalens, at the wavelengths from 440 nm to 700 nm with a step of 20 nm. The focal lengths of the bare lens monotonously increase as the incident wavelength changes from 440 nm to 700 nm, which represents the positive dispersion of the bare refractive lens. For comparison, the focal lengths of the hybrid metalens just fluctuate within a narrow range with the increasing wavelengths.

The detailed focal length shifts at different designed wavelengths are collected in figure 3(c). Compared with the maximum focal length shift of 0.834 mm for the bare lens, the corresponding merit of the designed hybrid metalens is 0.158 mm. The chromatic aberration is reduced by 81% with the metasurface. The residual chromatic aberration of the hybrid metalens is mainly caused by the linear approximation of the material index ( seen in Figure S2 (b) ) and the phase errors of the metasurface because of the limited meta-atom library. Furthermore, the focusing performance concerning the incident polarization is studied at the wavelengths of 640 nm. Figure 3(d) shows that the focusing efficiency nearly retains 75% as the polarization angle increases from 0° to 180°, which demonstrates the polarization insensitivity of the hybrid metalens.

## 3.2 Imaging with the achromatic hybrid metalens

Imaging characterization of the hybrid metalens is performed by commercial ray tracing software Zemax OpticStudio. The details can be found in Supplement 7.

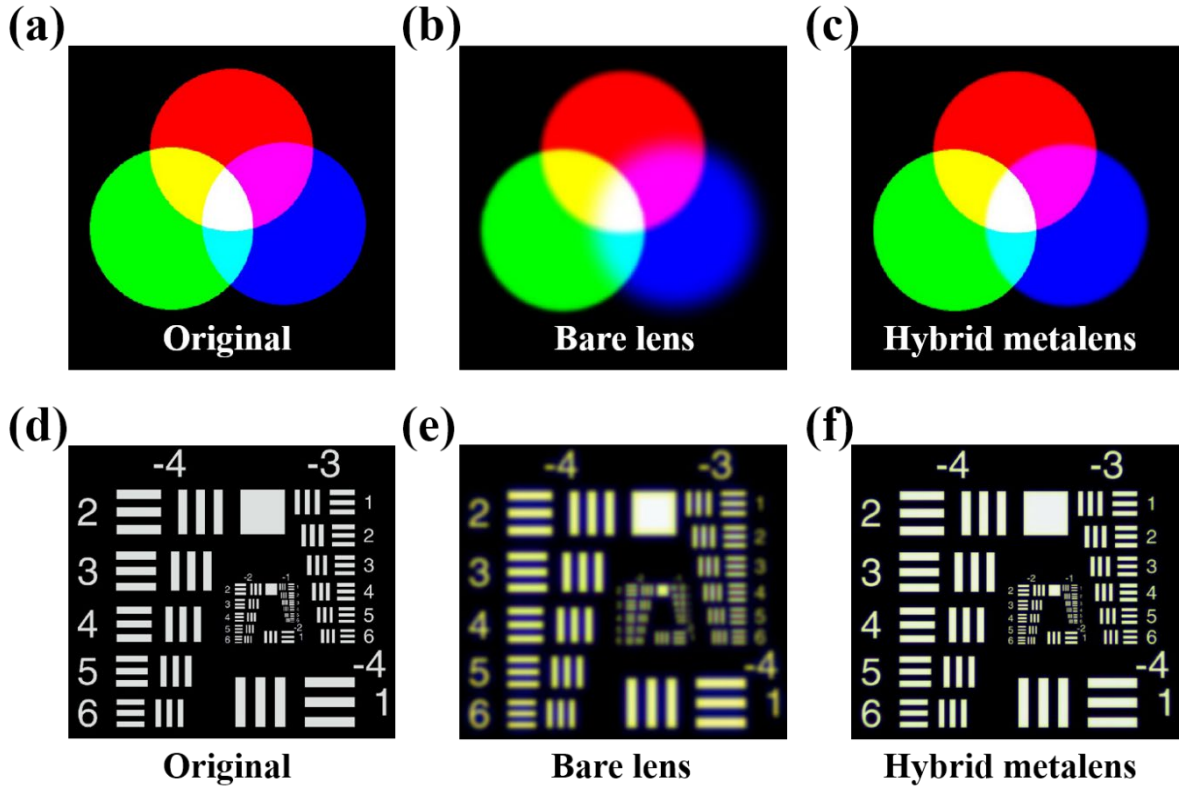

**Figure 4.** Imaging characterizations of the hybrid metalens. (a) Original pictures, simulated images with (b) the bare lens and (c) the hybrid metalens of the RGB circles. (d) Original pictures, simulated images with (e) the bare lens and (f) the hybrid metalens of the 1951 USAF resolution test chart. The object distance is 25 cm (the best distance of distinct vision for humans) and the height of the object is 10 mm.

The imaging simulation results of the RGB-circle image are plotted in Figures 4 (a), (b), and (c). Compared with the simulated image (Figure 4 (b)) of the bare lens, the simulated image of hybrid metalens has clear and sharp edges, especially the red and blue circles, owing to the great suppression of chromatism. The little Image blur in Figure 4 (c) results from the relatively small residual chromatic aberration. Simulated results of the 1951 USAF resolution test chart for the bare lens and the hybrid metalens are respectively shown in Figures 4 (e) and (f). The blurred and colored image is seen for the bare lens, while the hybrid metalens has a nearly white image with sharp edges illuminated under the broadband light source. The yellowish bars in Figure 4 (e) are because of chromatic aberrations. Specifically, lights with different wavelengths are focused on different image planes, resulting in colored rainbow-like halos. As depicted in Figure S8, the observed images turn different colors with varying image distances. Three wavelengths of light (606 nm, 535 nm, and 465 nm) are selected as incident wavelengths to imitate the broadband source in all imaging tests.

## 3.3. The accuracy of the semi-vector method

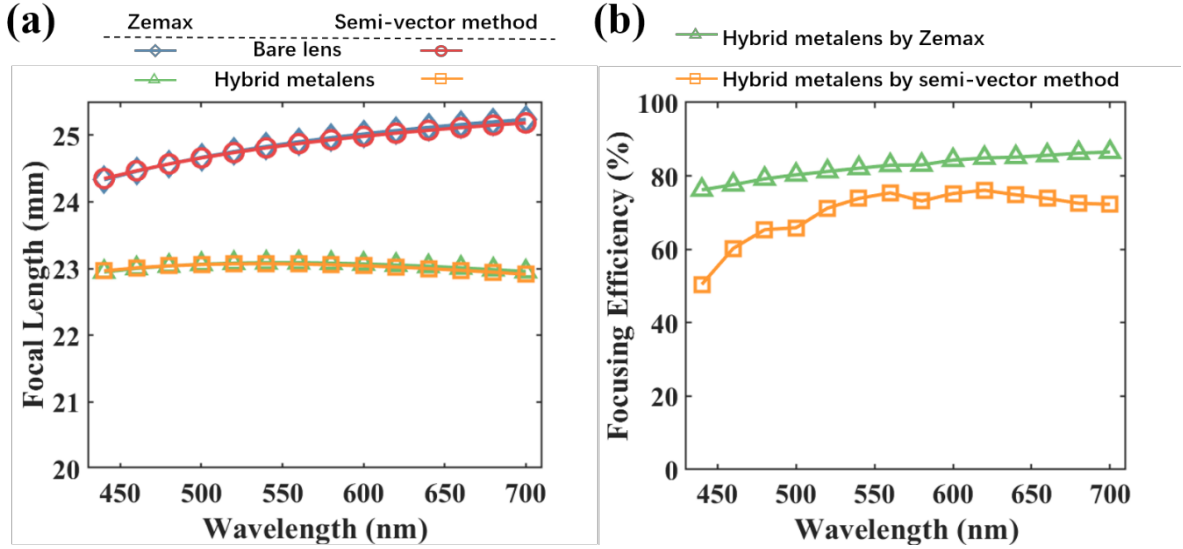

**Figure 5.** Comparison of the simulated (a) focal lengths and (b) focusing efficiency calculated by Zemax and the semi-vector method.

To prove the accuracy of the semi-vector method, the corresponding results calculated by Zemax are selected as the baseline. Figure 5 (a) shows the focal length of the bare lens and the hybrid lens separately calculated by ray tracing and the semi-vector methods. Taking the results of the bare lens as an example, the focal lengths calculated by Zemax OpticStudio agree well with those calculated by the semi-vector analysis method. The maximal deviation does not exceed 0.24%. There are similar results of the hybrid metalens. These slight deviations come from the different criteria of the focal length of the two methods. Specifically, the focal length in Zemax OpticStudio is the longitudinal position where the root mean square (RMS) of the wavefront aberration reaches the minimum, while the focal length in the semi-vector method is taken as the longitudinal position with the maximal central spot intensity.

Figure 5 (b) shows the focusing efficiency of the hybrid metalens calculated by the two methods. The focusing efficiencies calculated by the semi-vector method deviate from those calculated by Zemax OpticStudio. This deviation comes from the fact that the semi-vector method takes into consideration the complex diffraction effects and the nonuniform amplitude distributions of the meta-elements, while it is difficult for the software Zemax OpticStudio to simulate physical optical effects of meta-elements. These results verify the effectiveness and accuracy of the semi-vector method.

## 4. Conclusion

In summary, we have designed a centimeter-scale polarization-insensitive achromatic hybrid metalens in the visible, featuring a diameter of 1 cm and an F/# = 2.64. This hybrid metalens can reduce the chromatic aberration of the bare lens by 81% through the specially designed metasurface. Benefiting from the four-fold symmetry and wavelength-dependent phase response of the meta-atom, polarization insensitivity is achieved. Besides, the semi-vector method is proposed in this paper as a solution to perform the simulation of large-scale metasurface-refractive hybrid lenses. Noticeably, we generalize the design of polarization-insensitive metasurface correctors to suppress the chromatism of plano-convex lenses of any surface type, which are widely used in precision optical systems. The size of the hybrid metalens can be arbitrarily large in theory. In future work, the semi-vector method can be further extended to perform simulations of more complex optical systems.

**Acknowledgements**

This work is supported by the Natural Science Foundation of China (No.62075073, 62135004 and 62075129), the Fundamental Research Funds for the Central Universities (No. 2019kfyXKJC038), State Key Laboratory of Advanced Optical Communication Systems and Networks, Shanghai Jiao Tong University (No. 2021GZKF007), and Key R & D project of Hubei Province (No.12345678).

Tie Hu and Shengqi Wang contributed equally to this work.

**Author contributions**

T.Hu had the original idea, conceived the study, finished the FDTD simulations and ; S.Q.Wang helped the formula derivations and finished the Zemax simulations; T.Hu and S.Q.Wang performed the formula derivations with the help from X.Feng, Y.X. Wei; M.Zhao and Z.Y.Yang supervised the work and the manuscript writing. All authors discussed the results. T.Hu and S.Q.Wang cooperated to write the first draft of the manuscript, which was then refined by contributions from all authors.

# References

[1]     R. H. Katyl, *Appl. Opt.* **1972**, 11, 1241.
[2]     T. Stone, N. George, *Appl. Opt.* **1988**, 27, 2960.
[3]     M. Pan, Y. Fu, M. Zheng, H. Chen, Y. Zang, H. Duan, Q. Li, M. Qiu, Y. Hu, *Light: Science & Applications* **2022**, 11, 195.
[4]     L. Huang, S. Colburn, A. Zhan, A. Majumdar, *Advanced Photonics Research* **2022**, 3.
[5]     W. T. Chen, A. Y. Zhu, V. Sanjeev, M. Khorasaninejad, Z. Shi, E. Lee, F. Capasso, *Nat Nanotechnol* **2018**, 13, 220.
[6]     Z. Li, R. Pestourie, J. S. Park, Y. W. Huang, S. G. Johnson, F. Capasso, *Nat Commun* **2022**, 13, 2409.
[7]     O. Avayu, E. Almeida, Y. Prior, T. Ellenbogen, *Nat Commun* **2017**, 8, 14992.
[8]     F. Aieta, M. A. Kats, P. Genevet, F. Capasso, *Science* **2015**, 347, 1342.
[9]     E. Tseng, S. Colburn, J. Whitehead, L. Huang, S. H. Baek, A. Majumdar, F. Heide, *Nat Commun* **2021**, 12, 6493.
[10]    H. Liang, A. Martins, B.-H. V. Borges, J. Zhou, E. R. Martins, J. Li, T. F. Krauss, *Optica* **2019**, 6.
[11]    W. T. Chen, A. Y. Zhu, J. Sisler, Y. W. Huang, K. M. A. Yousef, E. Lee, C. W. Qiu, F. Capasso, *Nano Lett* **2018**, 18, 7801.
[12]    R. Sawant, D. Andrén, R. J. Martins, S. Khadir, R. Verre, M. Käll, P. Genevet, *Optica* **2021**, 8.
[13]    O. Kigner, M. Meem, B. Baker, S. Banerji, P. W. C. Hon, B. Sensale-Rodriguez, R. Menon, *Opt Lett* **2021**, 46, 4069.
[14]    F. Balli, M. Sultan, S. K. Lami, J. T. Hastings, *Nat Commun* **2020**, 11, 3892.
[15]    S. D. Campbell, D. Sell, R. P. Jenkins, E. B. Whiting, J. A. Fan, D. H. Werner, *Optical Materials Express* **2019**, 9.
[16]    Z. Lin, S. G. Johnson, *Opt Express* **2019**, 27, 32445.